\documentclass[11pt,twoside]{article}
\usepackage{asp2010}
\usepackage{natbib}

\aspvolume{471} 
\aspvoltitle{Origins of the Expanding Universe: 1912-1932}
\aspcpryear{2013} 
\aspvolauthor{Michael J. Way and Deidre Hunter, eds.}

\begin{document}

\title{Worlds and Systems in Early Modern Europe}
\author{Luc\'{\i}a~Ayala
\affil{Office for History of Science and Technology, University of California,\\
543 Stephens Hall \# 2350, Berkeley, CA, 94720, USA}}

\begin{abstract}
The structure, formation and evolution of the Universe were some of the main
topics in the scientific debates during the 17$^{\rm th}$ and 18$^{\rm th}$
centuries in Europe. They involved novel ideas on the cosmos, which concerned
aspects that were not considered before so emphatically, and which were
fundamental for the future development of astronomy. This paper presents a brief
account of several milestones within the gradual definition of pre-galactic
systems: the historical role of the tradition of the plurality of worlds, the
significance of Descartes, and the introduction of the Milky Way and nebulae in
the discourses around the cosmic structure.  \end{abstract}

\section{Introduction}
Before the gradual definition of galaxies that took place one century ago, the
formation, evolution, and structure of the Universe were put into question in
Europe, with a special intensity during the early modern period.\footnote{More
precisely from the late 16$^{\rm th}$ to the 18$^{\rm th}$ centuries.}
Many authors, even
without enough observational data of the most basic physical properties of
distant objects, attempted to give explanations to the nature and constitution
of the cosmos far beyond the Solar System.

One cannot expect to find in the 17$^{\rm th}$ and 18$^{\rm th}$ centuries
precise definitions of the characteristics of galaxies in terms that we now
understand. The lack of scientific information at that time makes this an
impossible task. Therefore, when reading astronomical texts from this period we
must be aware of the huge gap that separates us from them with regard to the
state of astronomy and, consequently, we cannot demand that they provide in this
context useful, measurable, or testable ideas for current astronomers.

Notwithstanding, the debates and hypotheses from this period fueled the further
development of science. Even the discourses that remained within a religious or
philosophical context triggered a field for discussion, pointing out topics and
research areas that were the grounds for future theories and observations. Some
authors may be considered from our perspective obscure thinkers or defenders of
erroneous pseudo-scientific statements; however, they were also significant,
insofar as they dealt with non-established -- and, sometimes, non-accepted --
ideas nevertheless containing key elements for setting the right (big) questions
astronomy had to answer.

Subjects from the 17$^{\rm th}$ and the 18$^{\rm th}$ centuries such as the
plurality of worlds, the vortex hypothesis, or the (in most of the cases highly
intuitive) ideas around the nature of the Milky Way and the nebulae, are usually
disregarded for not having significant scientific value. However, without
these debates, the formulation of `right' theories could not have taken
place.\footnote{For instance, even Slipher's initial redshift measurements at
Lowell Observatory in Flagstaff were motivated by some of these `unfortunate'
discussions. Let us remember that Percival Lowell (1855--1916) paid great
attention to the nebular hypothesis of planetary system formation and was a
fervent defender of the most controversial aspects of the plurality of worlds,
i.e. the existence of extraterrestrial life. See the text by Robert W. Smith in
this volume for an insight into this aspect of Slipher's work.\\Slipher is a
good example of the change of times, being placed between the long tradition of
speculative and non-testable hypotheses represented by Lowell, and the
definitive modernization of astronomy that was partially consolidated after his
measurements. For understanding the context of Lowell's ideas in this respect
see \cite{Crowe_1986}.}

In this text I present some basic considerations for a better understanding of
certain aspects from the history of astronomy in the early modern period that
made possible a gradual awareness of the actual structure of the Universe: the
vortex hypothesis, the plurality of worlds, and the first speculations on the
Milky Way and the nebulae, which were discussed in Europe by many scientists and
philosophers, such as Descartes, Swedenborg, or Kant, during the 17$^{\rm th}$
and 18$^{\rm th}$ centuries. Just after this period William Herschel focussed
the investigations on nebulae, finally leading to modern galaxy studies.

\section{The historical significance of the plurality of worlds}
One of the most radical starting points for the `new astronomy' that arose at
the end of the 16$^{\rm th}$ century and consolidated in the course of the
17$^{\rm th}$ and 18$^{\rm th}$ centuries in Europe was the idea that the cosmos
is composed of a plurality of worlds. This notion involved many innovative
ideas, which were widely discussed together or separately but, in any case,
which were perceived by the authors of that time as a single and solid
tradition. It generated a trend in scientific and philosophical texts that
contributed, on the one hand, to put emphasis on the question about the
structure of the Universe and, on the other, to popularize scientific topics in
an unprecedented way.

The Russian-French historian of science Alexandre Koyr\'e (1892--1964) proposed
in the middle of the 20$^{\rm th}$ century a very lucid approach to early modern
science based on the ideas around the structure of the Universe as a whole
\citep{Koyre_1957}. His inquiry was basically centered on how authors from the
17$^{\rm th}$ and 18$^{\rm th}$ centuries tackled the question of the infinitude
of the Universe. This way of approaching history, in which Giordano Bruno
(1548--1600) is more relevant and revolutionary than Copernicus (1473--1543),
puts forward important (and until then mostly forgotten) elements to comprehend
the change of mentalities that gradually led to the modern world. While Koyr\'e
attracted our attention to the radical transformation that took place during
this time with regard to the general notions on the structure of the Universe
(his famous statement ``from a closed world to an infinite Universe"), a
critical approach to the tradition of the plurality of worlds can provide the
most relevant clues for understanding how the particular constitution of this
new (infinite or, at least, unprecedentedly enormous) Universe was defined.

In the early 1980s the tradition of the plurality of worlds was subjected to two
major examinations that are still a reference for the field: the well-known
investigations by Steven J. \citet{Dick_1982} and Michael J. \citet{Crowe_1986}.
In spite of the significance of these texts in pointing out
the necessity of integrating this topic within a serious history of astronomy,
they both put the emphasis on one of the aspects regarding the plurality of
worlds, namely the possible existence of extraterrestrial life. However, even
though it was a very important topic in the history of astronomy, the most
groundbreaking element of this tradition concerned the specific notions around
the structure of the Universe, from which the possible existence of life in
other heavenly bodies was a side-effect, and not the core of the hypotheses.

The scientific, philosophical, and even literary tradition of the plurality of
worlds generated a general debate about the possible constitution of the cosmos
beyond the Solar System\footnote{The term \emph{world} was commonly used to
refer to a complete unit within the puzzle of the Universe, i.e. a system
containing (at least potentially) all the elements present in the Solar System:
a central star and several planets orbiting around it. These planets could have
related satellites and could be inhabited. As mentioned above, the main
references in the secondary literature about the plurality of worlds highlight
this last element, rather than giving a comprehensive approach to all the
implications that the worlds involve.} that stimulated much research,
conjecture, and passionate discussion on the topic throughout Europe.

From our perspective, at first glance one may be tempted to undervalue these
contributions for not being based on reliable scientific data. However, these
ideas can seem astonishing if we put it this way: in spite of the lack of
information, only starting from very simple observations and mainly proceeding
by reasoning, these authors managed to foresee that the Universe goes far beyond
the Solar System and the observable stars, that there are many autonomous and
interconnected systems similar to ours (as hypothesized in the first stages of
the plurality of worlds\footnote{See \cite{Fontenelle_1686}, one of the more
important texts that popularized this field.}), and that the Milky Way and the
nebulae are evidence for understanding the general structure (such as Wright of
Durham or Kant considered, at a time when the intense investigation of nebulae
had not yet started\footnote{When Kant published his \emph{Theory of Heaven},
William Herschel, the astronomer who would systematically look for and observe
nebulae, was only 17 years old. Nevertheless, it is important to remember that
Herschel was not the first person who noticed the existence of nebulae. Ancient
astronomers had observed many of them without the aid of a telescope. References
to nebulae appear also in other cultures. For instance, the very first image of
the Andromeda Nebula was published by the Arab astronomer al-Sufi in
\emph{Kitab al-suwar al-kawakib al-thabitha} [The Book of Fixed Stars] around
964. For a brief introduction to the main pre-Messier studies on nebulae see
\citet[][pp. 21-8]{Jones_1969}.}). Most of these authors were not able to give
precise (or accurate) explanations of the physical properties of the other
systems, nor to determine their size or their distances with respect to us. But,
in view of the state of the astronomical knowledge within this context at that
period, this is not a surprise. These early notions, though imprecise or
inaccurate compared to our current understanding, were nevertheless crucial in
establishing the framework of the field that had to be explored in the future.

The discussions on the plurality of worlds were not specific references
dispersed in various books. Instead, this set of ideas constituted a historical
tendency, a body of knowledge that put specific topics into question and
generated a particular tradition of authors and texts. As a matter of fact,
without taking this chapter of our history into serious consideration, it is
impossible to fully understand the early roots of modern notions on the
structure of the Universe.

\section{The leading role of Descartes' cosmology} A similar role to Giordano
Bruno in Koyr\'e's analysis is played by Ren\'e Descartes (1596--1650) within the history of
the plurality of worlds,\footnote{In the sense of being the main figure within
the history of a specific concept that modernized the picture of the Universe --
infinity in the case of Bruno, and the existence of many extrasolar systems in
Descartes'. Although both ideas were discussed since ancient times, Bruno and
Descartes initiated a new tradition making these concepts central in their
respective worldview.\\
One may think that Giordano Bruno would be the principal authority within the
plurality of worlds. However, he was not. Let us remember that the Inquisition
condemned him to death and burned him in Rome in 1600 for his heretical
thoughts. His misfortune made him a forbidden reference and, consequently, many
later authors of the plurality of worlds never cited Bruno. Instead, they
followed Descartes' physics -- be it for defending it, or for contradicting it.}
even though he did not focus on the `worlds' themselves, but rather on the
vortices.\footnote{We cannot forget that the vortices hypothesis had a parallel
and autonomous development with respect to the plurality of worlds. Descartes
defined the structure and evolution of the Universe in terms of vortices.
According to him, there were many -- literally, an ``indeterminate'' number of
-- systems similar to the Solar System, which were made of a fluid matter in
continuous motion. Owing to the fluidity and the intrinsic motion (hence the
name of vortices), the systems (and therefore the entire Universe) were in
constant change. They were autonomous entities but, at the same time, interacted
with the adjacent vortices, colliding with them, causing the extinction of
weaker or smaller systems, or increasing their size by merging with them.
Descartes explained the properties of the celestial matter as a basis to
understand the nature of cosmic evolution. However, he did not consider the
`worlds' as his followers did. For instance, the debate on extraterrestrial life
was irrelevant for him. For a comparison between the Cartesian vortices and the
picture of the Universe according to modern cosmology see
\cite{AyalaForero_2011}, also available online at
\texttt{http://arxiv.org/abs/1110.6772}.} With his physical model, Descartes
revolutionized the way of thinking about the structure of the Universe.
Furthermore, unlike Bruno, he was not prosecuted for his thinking, such that
this kind of debate became generally accepted.

In his major scientific texts \cite{Descartes_1644,Descartes_1664}
developed the first serious attempt in early modern Europe to understand the
nature and physics of extrasolar systems. He postulated the probable existence
of many entities outside the Solar System, which should be similar to it in
their basic characteristics but, at the same time, independent, autonomous, and
able to interact among each other. 

Descartes not only described a Universe made up of independent and distant
systems, but also how these vortices were formed and were subjected to an
ongoing evolution. Thus, the works mentioned above were entirely devoted to
explaining the formation, evolution and construction of the Universe. During the
centuries that preceded him, the origin of the Universe was sometimes discussed,
but mostly within the religious framework of the Bible (more precisely the Book
of Genesis). Through his works, Descartes initiated the trend of taking
cosmology into consideration strictly within a scientific perspective.
Afterward, other authors tackled the origin and evolution of the Universe as a
normal topic in astronomy, while there were still authors who confused them with
religious notions. This was not the case with Descartes; he kept religious
explanations out of his understanding of the physical world.

Fascinated by the idea of a vast Universe composed of a multitude of systems,
many authors embraced the vortices as the most revolutionary hypothesis in
astronomy so far, which became the paradigm for a secularized
science.\footnote{Because his explanations were independent from religious
thought, many authors from the French Enlightenment defended Descartes and the
vortices, at a time when their scientific truthfulness was seriously
questioned.} This new corpus of ideas achieved exceptional popularity mainly
owing to the great success of the book \emph{Conversations on the Plurality of
Worlds}, published by Bernard Le Bovier de Fontenelle (1657--1757) in Paris in
1686 and republished again and again in many countries and languages throughout
the course of several centuries \citep[see][]{Fontenelle_1686, Ayala_2011}.
Fontenelle, following Descartes' ideas, helped in a significant way to change
the general opinion of Europeans regarding the structure of the Universe.

At that point in history, neither the Milky Way nor the nebulae had a prominent
position in the explanations on the structure of the Universe. Instead, the
Solar System -- or the solar vortex, which according to the Cartesian
terminology, contained the Solar System -- was the paradigmatic piece in the
constitution of the cosmos. In other words, in the pre-galactic cosmos, the
Solar System was the referential unit. Although the Cartesian vortices and the
worlds were not defined according to the dimensions or properties of the
galaxies, they represented a necessary initial step within a gradual change of
paradigms regarding the constitution of the Universe.\footnote{These research
areas motivated a growing interest in the structure of the Universe -- a
Universe that was gradually conceived as larger and more complex with the
passing of time, and went from being based on vague suppositions to being
confirmed by observations and measurements.}

\section{Swedenborg's cosmos: ``Expansion" in an ``active space"} One century
after the publication of Descartes' texts, his vortex hypothesis was still
considered a valid model. This was in spite of the supporters of Newton, who
usually set them in opposition and, as a result, systematically discarded all
ideas coming from the French scientist. One of the late advocates of the
vortices was the Swedish scientist and theologian Emanuel Swedenborg
(1688--1772), a `mystic scientist' in a similar way to Thomas Wright of Durham
(in fact they were contemporaries; see Sect. \ref{MW}), as he was also driven by
scientific and religious motivations without making any distinction between
them. Nevertheless, the most analytical and scientific part of their hypothesis
was relevant in view of the later development of science, especially in the
context of cosmology.

In \emph{The First Principles of Natural Things}
\citep{Swedenborg_1743},\footnote{For the quotes from this text I use an English
translation done one century later. Since the English of the translation is as
obscure as the original Latin, sometimes I add an explanation within the quote
itself in order to make it easier to understand. The original edition was the
first volume of \emph{Opera philosophica et mineralia}
\citep{Swedenborg_1743}.} Swedenborg developed a complete description of the
formation and evolution of the Universe according to the vortices, while
providing many of the details that had remained unelaborated by Descartes. He
described the Universe in terms of its transformation processes: the many
systems are (after Descartes) in a constant process of collision, merging,
vanishing, and change. The systems of his cosmos were being created or destroyed
at all times. He conceived the cosmic structure as a Russian doll, in which all
elements were contained within larger structures:

\begin{quote}
Hence may arise new heavens one after the other; in these heavens, new vortices
and mundane [i.e. planetary] systems; in these vortices and systems, new
planets; around the planets, new satellites \citep[Vol.\ 2][p.\
240]{Swedenborg_1846}.  \end{quote}

Swedenborg identified two notions from which to analyze the cosmos: its ``figure
and motion or, geometry and mechanism" \citep[Vol.\ 2][p.\ 261]{Swedenborg_1846}
-- what we would now call its `structure' and `evolution.' Following these
parameters, he explained the process of the formation of the Solar System, which
took place within the solar vortex. As stated by the vortex hypothesis, the
particles are always in motion. According to that, Swedenborg defined the
Universe as an ``active space" many times.

In the beginning, all matter in the Solar System was condensed around the Sun,
like a crust. By the action of its intrinsic motion, this matter separated from
the center (i.e. the proto-Sun) and expanded itself until it reached the borders
of the solar vortex:

\begin{quote}
[This crust of matter], being endowed with a continual gyratory motion round the
sun, in the course of time removes itself farther and farther from the active
space [i.e. the Sun at the center]; and, in so removing itself, occupies a
larger circle of space \citep[Vol.\ 2][p.\ 261]{Swedenborg_1846}.  \end{quote}

When this matter reaches the border of the vortex, it collapses and forms a kind
of belt that is still rotating round the Sun. Since the motion does not cease,
the belt or ring naturally continues evolving: it condensates and forms planets
and satellites. After the system has been formed, Swedenborg considered
different possibilities for its later evolution: it could subside inwardly or
outwardly toward another system but, in general, the proto-solar system
should reach equilibrium
with the volume of the vortex.\footnote{Let us note that for Swedenborg, the
``vortices" are like a container where the ``systems" are formed. The
explanation of Swedenborg was echoed (via Kant) by Pierre-Simon Laplace
(1749--1827) in the famous Kant-Laplace nebular hypothesis for planetary system
formation; see \cite{Laplace_1796}. Nevertheless, Laplace did not mention
Swedenborg as a source, and exclusively mentioned Georges Louis Leclerc, Comte
of Buffon (1707--1788), as the only person who had seriously dealt with
cosmology so far. Contrary to Swedenborg, Laplace put more emphasis on the role
of gravity in the formation process.} For the Swedish scientist this process of
consecutive transformations (first, expansion in all directions; second,
formation into a ring; third, transformation into ``larger and smaller globes",
that is, the heavenly bodies) was a scientific fact:

\begin{quote}
That the expanse becomes attenuated in consequence of forming a larger circle,
is a purely geometrical fact \citep[Vol.\ 2][p.\ 262]{Swedenborg_1846}.
\end{quote}

The expression ``active space" is continually used to refer to the Universe or
parts of it. As well, the noun ``expanse" was also used frequently to describe
the process taking place in the formation of the Solar System, such as in this
case:

\begin{quote}
That this belt, which is formed by the collapse of the crustaceous expanse [i.e.
the expansion of the crust], gyrates in a similar manner; removes itself to a
farther distance; and by its removal becomes attenuated till it bursts, and
forms into larger and smaller globes; that is to say, forms planets and
satellites of various dimensions \citep[Vol.\ 2][p.\ 262]{Swedenborg_1846}.
\end{quote}

Swedenborg did not conceive of the expansion of the Universe in the same terms
as we understand it today. Nevertheless, he put emphasis on a key consequence of
the physical model based on vortices that was not expressed so clearly before:
the expansion of the systems as a natural process of their evolution. From a
dense, small area, by their intrinsic spinning motion, the systems progressively
become larger:

\begin{quote}
The tendency of the crust is to fly off [the center] to a greater distance
\citep[Vol.\ 2][p.\ 262]{Swedenborg_1846}.  \end{quote}

Like other authors under the influence of the plurality of worlds, Swedenborg
understood that the Solar System was the essential piece in the construction of
the Universe. For him, describing the process of the formation of the Solar
System implied giving an explanation of the formation of planetary systems in
general, which concerned the cosmic structure as a whole. Therefore, the
implications of his description of the formation of the Solar System went beyond
the system itself, which was assumed as a model to be applied to the general
structure. Insofar the authors of this period used to operate by analogy
(explaining the observed phenomena of the Solar System in terms of generalizable
phenomena), the cosmogony of many authors, like Swedenborg, can be understood as
general cosmology -- at least, it was so for them.

\section{The Milky Way and the nebulae come into play}
\label{MW}
When in 1750 Thomas Wright of Durham (1711--1786) published \emph{An Original
Theory or New Hypothesis of the Universe},\footnote{For a comprehensive analysis
of Wright's hypothesis see the reprint and introduction by Michael A. Hoskin
\citep{Wright_1971}.} the idea of the plurality of worlds had already been
established and validated by a century of intense discussions, a number of
publications and diverse reactions. 

Wright's principal goal was to integrate the new structure of the Universe into
a religious worldview. The English astronomer applied to the usual picture of
the cosmos in the 17$^{\rm th}$ and 18$^{\rm th}$ centuries a personal (thus the
``original" included in the title of the book), ambitious and obscure religious
order in which God and moral values acquired the same status in the physical
world as the stars and the planets. According to him, the Universe is composed
of an infinite number of autonomous entities similar to our System, which,
following the common notions from the plurality of worlds, can be planetary or
star systems that can contain (inhabited) planets or only stars:

\begin{quote}
[\ldots] the visible Creation is supposed to be full of siderial [sic] Systems
and planetary Worlds, so on, in like similar Manner, the endless Immensity is an
unlimited Plenum of Creations not unlike the known
Universe \citep[p.\ 83]{Wright_1750}.  \end{quote}

Despite the fact that the religious aspect was fundamental to his thoughts and
encompassed his main concerns, it is important to mention him in this context
for one reason: the role he conferred to the Milky Way and the nebulae as
starting points for shaping a hypothesis about the structure of the Universe.

He asserted that ``there are other luminous Spaces in the starry Regions, not
unlike the Milky Way" \citep[p.\ 42]{Wright_1750}. After explaining his bizarre
ideas of spheres within spheres that have God as the common center of the
heavenly bodies in each system (or ``Creation"), he argued that the nebulae
observed in the sky prove his ideas, since they are actually the distant systems
he describes:

\begin{quote}
That this [the Universe structure he proposed] in all Probability may be the
real Case, is in some Degree made evident by the many cloudy Spots, just
perceivable by us, as far without our starry Regions, in which tho' visibly
luminous Spaces, no one Star or particular constituent Body can possibly be
distinguished; those in all likelyhood [sic] may be external Creation, bordering
upon the known one, too remote for even our Telescopes to
reach  \citep[pp.\ 83-4]{Wright_1750}.  \end{quote}

Regarding the Milky Way, Wright stated:

\begin{quote}
[H]ow far I have succeeded in my designed Solution of the \emph{Via Lactea},
upon which the Theory of the Whole is formed, is a Thing will hardly be known in
the present Century, as in all Probability it may require some Ages of
Observation to discover the Truth of it \citep[p.\ 66]{Wright_1750}.
\end{quote}

According to him, the ``Milky Way" we see crossing the sky is the product of a
visual effect. For him it has in fact a spherical shape, and is like a shell.
The Solar System is embedded in it; thus, we observe the shell of the Milky Way
through a tangent view from within. Although this reasoning does not really
work, he saw in it a proof for the impression that the system is the shape of a
plane. Hence, his famous description of the Milky Way was presented as a
supposition, since for him the plane is a visual effect and not its actual
shape:

\begin{quote}
Let us imagine a vast infinite Gulph, or Medium, every Way extended like a
Plane, and inclosed between two Surfaces, nearly even on both
Sides \citep[p.\ 62]{Wright_1750}.  \end{quote}

In other words, within his system our Galaxy was a sphere, but we perceive it as
a plane because we are immersed in the middle of its shell. Consequently, ``to
an Eye situated near the Center Point," the Galaxy appears to be a ``Zone of
Light."

He insisted on the fact that this explanation concerns only the visible aspect
of the Universe that is accessible to our senses (i.e. as an optical effect),
but is not its real structure. He presented this hypothesis as a possible
explanation of what we see in the sky but, as he asserted, ``I don't mean to
affirm that it really is so in Fact" \citep[p.\ 62]{Wright_1750}. For him, what
was really important was to reconcile the real structure of the cosmos with
``the visible Order of its Parts," in this case, the observed Milky Way. This
task was obviously hard to fulfill, since his cosmic model was founded on false
grounds and was extremely complex due to mixing physical phenomena with
religious elements.

Wright failed to outline a correct (and even reasonable) structure of the
Universe, but he did foresee that the Solar System, as traditionally argued in
the plurality of worlds by then, was not the proper unit to consider regarding
the general scale; instead, it was an element of a larger structure composed of
systems with a much bigger size. These aspects of his hypothesis (or `theory'
according to his own terminology), although being secondary elements within his
worldview, acquire relevance in view of the future development of astronomy.

Only five years after the publication of \emph{An Original Theory} by Wright,
and at the age of 31, Immanuel Kant (1724--1804) brought out the
\emph{Allgemeine Naturgeschichte und Theorie des Himmels} [Universal Natural
History and Theory of Heaven] \citep{Kant_1755}.

Contrary to popular opinion, the term `island universes' never appears as such
in the \emph{Theorie des Himmels}, although the idea behind it has a notable
presence. Kant did discuss other systems that are similar to the Milky Way, and
he referred to them using
expressions like ``similar systems" (\"ahnliche Systeme), ``Systemata", ``world systems"
(Weltgeb\"auden, Weltsystemen), ``world orders" (Welt\-ordnungen), ``heavenly hosts"\newline
(Himmelsheere), or ``solar systems"
(Sternensysteme). The word `island' (Insel in German) appears a couple of times
in this book, but to refer to a real island (Jamaica), or as a metaphor applied
to planets without inhabitants.\footnote{He mentioned that the regions of the
sky without any inhabited planet are like a ``desert or island" (``eine W\"uste,
oder Insel") in the ocean of the Universe \citep[p.\ 175]{Kant_1755}.}

By the middle of the 18$^{\rm th}$ century, when Kant published this book, the
concept of the plurality of worlds was already well-established and was a common
topic of discussion in both specialized and popular literature. In fact, he used
the same expressions (see the terms listed above) that were usual in this field
to designate the structural components of the Universe. But Kant went further
and gave the best description (in a contemporary sense) of the cosmic structure
so far.\footnote{Contrary to Wright, Kant expressed his ideas very clearly
without metaphors or obscure analogies, and exclusively pursued scientific
goals.}

Instead of talking about the concept of island universes, Kant defined what he
called a ``systemic constitution of the Universe."\footnote{Literally:
``systemische Verfassung des Weltbaues."} In essence, it involved the definition
of the components of the Universe's structure as `systems': the Solar
\emph{System} was the basic element; it belonged to the category of planetary
\emph{systems}, which were associated with the stars seen in the sky (each
star was considered to be a sun, in the sense of forming a planetary system); a
group of many planetary \emph{systems}, including ours, forms the Milky Way,
which was also defined as a \emph{system} (and described as a plane that
constitutes an area of relationships -- or ``relational area" -- caused by its
gravitational forces, in which all elements are rotating round its center and
whose general shape is elliptic); finally, the nebulae are also \emph{systems}
with similar characteristics to the Milky Way and which are so far away from us
that their constitutive heavenly bodies blur themselves in a misty appearance.
Therefore, Kant's worldview was an intermediary step between the plurality of
worlds and more modern day conceptions.

He defined our Galaxy as follows:

\begin{quote}
The shape of the heavens of the fixed stars has no other principle than the same
system constitution in large that the planetary world-edifice [i.e. the Solar
System] has in small, inasmuch as all suns form a system whose general
relational plane is the Milky Way \citep[p.\ 8]{Kant_1755}.\footnote{``Die
Gestalt des Himmels der Fixsterne hat also keine andere Ursache, als eben eine
dergleichen systematische Verfassung im Grossen, als der planetische Weltbau im
Kleinen hat, indem alle Sonnen ein System ausmachen, dessen allgemeine
Beziehungsfl\"ache die Milchstrasse ist."\\Note: I prefer to give my own
translation of Kant's quotes, since some important concepts are missed or not
satisfactorily interpreted in the official translations I have consulted.}
\end{quote}

This quote summarizes the detailed description of the structural components of
the Universe expounded in the first part of his book. He built his hypothesis on
several steps woven according to increasing scales.

First, he considered that the structure of the Solar System has a similarity
with the structure of other neighboring systems: \begin{quote}
[\ldots] the fixed stars, just like many suns, are the center of similar
systems, all of which may be arranged just as big and as ordered as in our
system \citep[p.\ 2]{Kant_1755}.\footnote{``[\ldots] die Fixsterne als eben so
viel Sonnen, Mittelpunkte von \"ahnlichen Systemen seyn, in welchen alles eben
so gro{\ss} und eben so ordentlich als in den unsrigen eingerichtet seyn mag."}
\end{quote}

Second, he stated that all together, they belong to the same entity, namely the
Milky Way. For Kant the Milky Way is a big system composed of many star or
planetary systems, all with a similar basic structure and characteristics but of
a different size and with different elements. He recognized that our Galaxy
generates an ``area of relationships" which, by virtue of gravitational forces,
bunch all the galactic celestial bodies and systems together. He explained how
the gravity generated by the Galaxy acts upon the systems belonging to it in a
similar way such that the gravity produced by the Sun acts upon the celestial
bodies of the Solar System, and likewise the gravity of the other stars has an
effect on the respective planetary systems that they form.

In a third step, he followed Wright's hypothesis that nebulae are systems that
are similar to the Milky Way. After explaining some of the characteristics of
nebulae, he convincingly asserted that:

\begin{quote}
All this agrees so completely that we should consider these elliptic figures
[i.e. the nebulae] as similar world orders, so to speak, Milky Ways \citep[pp.\
14-5]{Kant_1755}.\footnote{``Alles stimmet vollkommen \"uberein, diese
elliptische Figuren vor eben dergleichen Weltordnungen, und so zu reden,
Milchstrassen zu halten."\\Kant followed the French astronomer Pierre-Louis
Moreau de Maupertuis (1698--1759), who observed that many nebulae are not a
perfect circle or sphere, but that they indeed have an elliptic shape
\citep{Maupertuis_1732}. Kant was absolutely convinced of this and, as he
usually operated by analogy, concluded that the Milky Way must also be an
elliptic disc.} \end{quote}

The German philosopher took a step further than Wright and asserted that the
plane ascribed to the Milky Way was not a mere optical effect far from reality,
but a hint of its actual shape. According to that, its appearance in the sky is
in fact due to its actual shape as a disk, and because the Solar System, from
which we observe, is embedded in its plane. To him, this is an ``undoubted
phenomenon" \citep[p.\ 4]{Kant_1755}.

Furthermore, Kant linked (again by analogy) the rotating motion of the Solar
System to the entire Galaxy. He even brought the appearance of new stars into
question, asking himself if it might be the case that the rotation of the Galaxy
causes certain heavenly bodies to become visible or invisible to us depending on
their position at each moment.

To Kant, this description of the Universe's structure is nothing but obvious.
For instance, he was astonished by the fact that earlier astronomers did not
realize the shape of the Milky Way from its appearance in the sky, which is so
clearly distinguishable:

\begin{quote}
It is surprising that the observers of the sky had not long ago been moved by
the nature of this zone clearly distinguishable in the sky, to determine from it
the peculiar position of the stars [that are part of it] \citep[p.\
3]{Kant_1755}.\footnote{``Es ist zu bewundern, da{\ss} die Beobachter des
Himmels durch die Beschaffenheit dieser am Himmel kenntlich unterschiedenen Zone
nicht l\"angst bewogen worden, sonderbare Bestimmungen in der Lage der Fixsterne
daraus abzunehmen."} \end{quote}

\section{Conclusion}
The structure, formation and evolution of the Universe were intensively
discussed during the early modern period in Europe, becoming one of the most
fascinating topics to affect the status of astronomy -- a field in radical
transformation at that time. On the one hand, the hypotheses presented above run
parallel to the development of other aspects from the history of astronomy,
mainly represented by Newton's ideas and his followers. On the other hand, they
were the basis for future investigations, such as William Herschel's interest in
nebulae.\footnote{For more information on the later evolution of the
investigations of nebulae see \cite{Hoskin_2012,Jones_1969,Steinicke_2010}.}
These hypotheses, assumptions, and conjectures make possible a general awareness
of the existence of many systems outside the observable and near cosmos. The
curiosity, questioning, and revolutionary approaches that arose after them are
an essential part of the evolution of astronomy in the context of the structure
of the Universe, but also of the evolution of mentalities and philosophical
paradigms that led to the modern worldview.


\end{document}